# Fast Proteome Identification and Quantification from Data-Dependent Acquisition - Tandem Mass Spectrometry using Free Software Tools

Jesse G. Meyer

Department of Chemistry; Department of Biomolecular Chemistry; National Center for Quantitative Biology of Complex Systems; University of Wisconsin – Madison, Madison, WI, 53706.
* Correspondence: jessegmeyer at gmail dot com

**Abstract:** Identification of nearly all proteins in a system using data-dependent acquisition (DDA) mass spectrometry has become routine for simple organisms, such as bacteria and yeast. Still, quantification of the identified proteins may be a complex process and require multiple different software packages. This protocol describes identification and label-free quantification of proteins from bottom-up proteomics experiments. This method can be used to quantify all the detectable proteins in any DDA dataset collected with high-resolution precursor scans. This protocol may be used to quantify proteome remodeling in response to a drug treatment or a gene knockout. Notably, the method uses the latest and fastest freely-available software, and the entire protocol can be completed in a few hours with data from organisms with relatively small genomes, such as yeast or bacteria.

## 1. Introduction

Tandem mass spectrometry is currently the best method for unbiased, high throughput protein identification[1]. In fact, the entire yeast proteome can be routinely quantified in under one hour [2,3]. Still, the quantification of proteome remodeling can be a slow and difficult process, and many options are available for the multiple steps of analysis [4–6]. The main aim of this protocol is to identify and quantify proteins starting from raw mass spectrometry data. This protocol can be applied to data for any type of biological study, such as diseased and healthy tissue. This protocol combines the newest software tools to achieve the quantitative result as quickly as possible. All the tools described in this protocol are freely available and adaptable to different types of workflows, such as isotope labeling [7].

## 2. Experimental Design

This protocol describes data analysis only, as there are many other examples of protocols for data collection [3]. Alternatively, data from a previously-published study can be downloaded from a public repository for re-analysis. Starting with the raw mass spectrometry data, this protocol describes all analysis steps for peptide and protein identification, quantification, and statistical testing. The method uses the GUI for MS-Fragger to identify proteins by database searching [8], PeptideProphet and ProteinProphet to refine those identifications [9,10], Skyline to perform quantification [11], and MSstats to perform statistical testing [12]. Researchers planning proteomics experiments who plan to use this protocol should collect biological replicates of their controls and perturbation of interest. The sensitivity of detecting protein changes will depend greatly on the number of replicates collected and the variability to the data. This protocol should yield clear changes when used for quantification from significant perturbations, such as drug treatments. The tutorial data is from a previous study looking at single-gene knockouts in yeast [13], and is available from massive.ucsd.edu under the accession MSV000083136 (ftp://massive.ucsd.edu/MSV000083136/raw/). Scheme 1 summarizes the experimental design including the time needed to complete every stage.

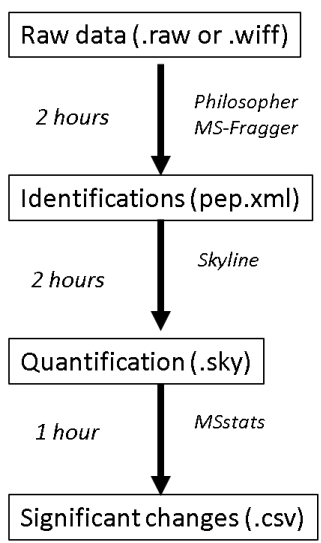

**Scheme 1.** Overview of the protocol steps, software used, and time required for each step.

*2.1. Materials*

- Raw mass spectrometry data from data-dependent acquisition proteomics experiment (tutorial data available from: ftp://massive.ucsd.edu/MSV000083136/raw/)
- MSconvert Software (http://proteowizard.sourceforge.net/download.html)
- FragPipe Software v7.1 (https://github.com/chhh/FragPipe/releases/tag/v7.1)
- MS-Fragger Software (https://bit.ly/2z6dzXa)
- Philosopher Software (https://github.com/prvst/philosopher/releases/tag/20180924)
- Skyline Software (https://skyline.ms/project/home/software/Skyline/begin.view)

*2.2. Equipment*

- 64-bit computer with at least 8 GB of RAM and at least quad-core i5 processor or equivalent

**3. Procedure**

*3.1. Install Required Software. Time for Completion: 1 Hour.*

Follow the instructions on the developer's websites to install the software programs described under section 2.1.

*3.2. Identify peptides by database searching. Time for Completion: 2 Hours.*

*3.2.1.   Convert Raw Mass Spectrometry Data to mzXML.*

1. Navigate to your system folder containing the raw mass spectrometry data.
2. Select your raw data files (.raw files from Thermo instruments, .wiff files from ABsciex instruments).
3. Right click on the selected files, and choose "Open with MSconvertGUI"
4. In the options box below the output directory, adjust the settings to output format = "mzXML", Binary encoding precision= "64-bit", and check the boxes next to "write index", "Use zlib compression", and "TPP compatibility".
5. In the filter box, select the dropdown box, and choose "Peak Picking". Do not change the settings that pop up, and click "Add". Your window should look like Figure 1.
6. At the bottom right corner, click "Start", and wait for your files to finish converting to mzXML.

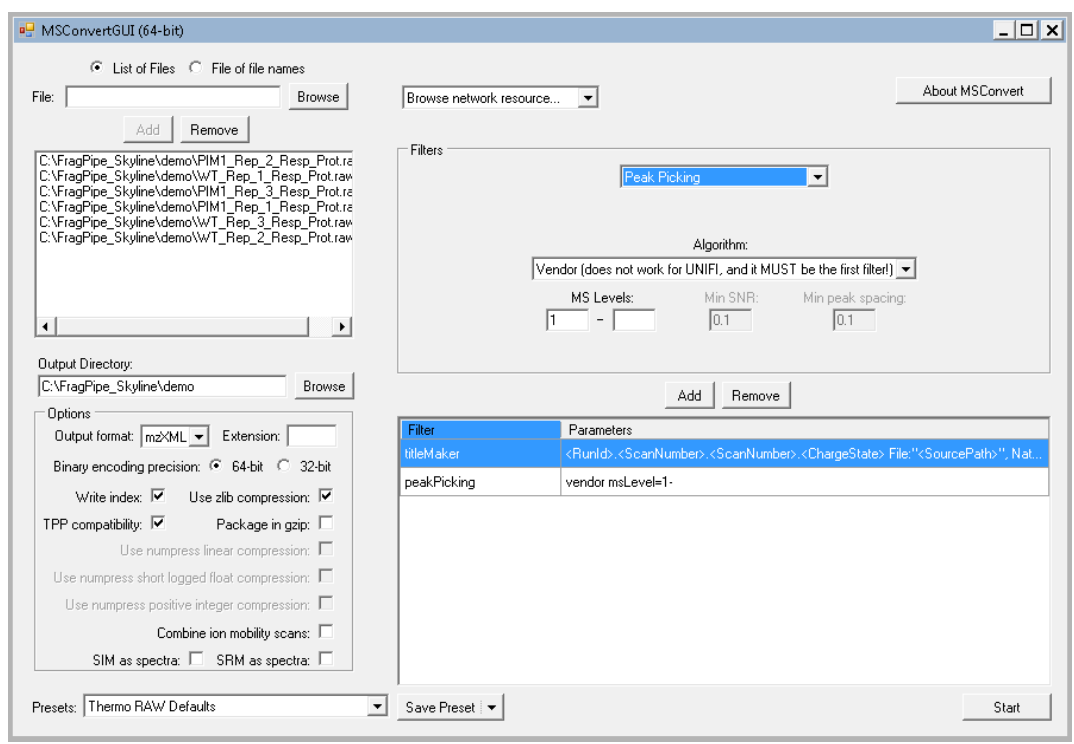

**Figure 1.** Settings used for MSConvertGUI to convert data from Orbitrap instruments.

*3.2.2.   Prepare the organism-specific database using philosopher.*

1. Navigate to the list of uniprot proteomes in your web browser: https://www.uniprot.org/proteomes/

2. Type the name of your organism in the search box. With the tutorial data, the data is from *Saccharomyces cerevisiae* (UP000002311).
3. Copy the uniprot ID from the column to the left of its name.
4. Open a windows command prompt (click the "start" button on the lower left corner, type "cmd" and hit enter.
5. Navigate to the location of your philosopher executable using the command "cd [full path to folder]" (Figure 3). For this tutorial, we created a file on the C:\ drive with the executables, so we use:

    cd C:\FragPipe_Skyline\
6. Initialize your philosopher workspace by typing (Figure 2):

    philosopher_windows_amd64.exe workspace –init

    Where the first command is thename of your philosopher executable.
7. Download your organism database and add contaminants and decoys by typing (Figure 2):

    philosopher_windows_amd64.exe database --prefix rev_ --contam --id UP000002311

**Figure 2.** Initialize a philosopher workspace and automatically download and format a protein sequence database.

*3.2.3. Peptide identificaiton by database search using FragPipe interface to MSFragger.*

1. Open FragPipe by double clicking on Fragpipe.bat
2. The FragPipe window should pop up and prompt you for locations of the MSFragger.jar and philosopher.exe. Click browse to navigate to their locations, or click the download buttons for links to their download locations (Figure A1).
3. Select the second tab "Select LC/MS Files", and add the .mzXML files we created in step 2 with either dragging and dropping them into the large white box, or by clicking "Add files" and navigating to their location (Figure A2).
4. Select the third tab "Sequence DB", and add the FASTA file we created in step 3 by clicking the "Browse" button and navigating to its location (Figure A3).
5. Select the fourth tab "MSFragger", and click the button on the top left "Defaults Closed Search". Two boxes will pop up asking to confirm. Click "yes" on both boxes.
6. Change the precursor and fragment mass tolerances to values that reflect your instrument performance. For the tutorial data, the precursor tolerance we'll use is 10 ppm, and fragmentation spectra were collected at low resolution in the ion trap, so from the dropdown box to the right of "fragment mass tolerance" set the value to "ABS" and enter 0.35 (Figure A4).
7. Leave the remaining tabs with default settings, and select the last tab "Run". Set your output file location by clicking the "Browse" button, and then click "Run" to start the database searches, PeptideProphet, and ProteinProphet analysis.

*3.3. Quantify peptides with Skyline. Time for Completion: 2 Hours.*

1. Open Skyline by clicking the windows start button, typing "Skyline", selecting skyline, and hitting enter.
2. On the Startup page, click the option in the top middle, "Import DDA Peptide Search".
3. Skyline will prompt you to save the document. Save the document, and then Skyline will prompt you with the "Import Peptide Search" box. Set the cutoff score to 0.99, and then click "Add Files…" and navigate to your MSFragger output folder. Select the files starting with "interact-" and ending with "pep.xml" (Figure B1). Click "Next" and skyline will start reading the files and building your spectral library.
4. Skyline will then prompt you to extract chromatograms, and should find your .mzXML files. If not, browse to add them (Figure B2).
5. Skyline will prompt you to optionally remove any common prefix from the file names, click "remove" and then will prompt you to add modifications it found in your database search results. Select the modifications you expect and want to use for quantification, in our case N-terminal acetylation. Click "Next".
6. Skyline will prompt you to configure the full-scan settings used for signal extraction. For our tutorial data, set the precursor charges to "2,3,4,5", and leave the other defaults unchanged (Figure B3). Click "next".
7. Skyline will prompt you for the database used to search for peptides, the enzyme used to digest to proteins, and the number of missed cleavages allowed. Leave the Enzyme as "Trypsin", the missed cleavages as 5, and click "browse" to navigate to the FASTA file created in step 3 (Figure B4). Click "Finish" and Skyline will begin adding the proteins that match identified peptides.
8. Skyline will then prompt you about what proteins you want to keep. You can filter based on the number of proteins identified and whether you will allow duplicate peptides. For the tutorial data, keep the default of 1 peptide per protein, and check the box next to "Remove duplicate peptides" (Figure B5). Skyline will then begin extracting the precursor peaks for the identified peptides. You can proceed with the next steps while Skyline continues to import the raw data. The raw data import will take about 1 hour depending on the speed of your computer.

*3.4. Statistical testing with MSstats: Time for Completion: 1 hour.*

1. Install MSstats within Skyline by going to the "Tools" menu > "Tool Store", and then selecting MSstats from the list along the left side and clicking "install". The installer will also install R, and may take a few minutes.
2. Go to the "Settings" menu > "Document Settings". Check the boxes next to "Condition" and "BioReplicate" and click OK.

3. Click on the "View" menu and select "Document Grid". In the Document Grid popup box, click on the "Views" dropdown menu and select "Replicates". Under the "Condition" column, select "Disease" for the PIM1 replicates, and "Healthy" for the WT replicates. Under the BioReplicate column, assign the number of biological replicate to each sample (Figure B6). Close the Document Grid box.
4. Once the data has finished importing, go to the menu "Tools"> "MS stats" > "Group comparisons". Skyline will take a moment to write a report file for input to MSstats. In the popup box "MSstats Group Comparison", name the comparison and leave the other settings as default, then click OK (Figure B7). The "Immediate Window" will pop up and display the status of the process.
5. After the "Immediate Window" displays "finished", the MSstats output will appear in the same directory as the Skyline file.
6. Skyline can be used to directly inspect the changes reported by MSstats. To arrange your Skyline workspace for easy data inspection, go to the "View" menu > "Arrange Graphs" > "Tiled". Also add the peak area comparison window by selecting "View" > "Peak Areas" > "Replicate Comparison". Drag your "Peak Areas – Replicate Comparison" window to the bottom of the master Skyline window and drop it over the down arrow that appears to anchor it at the bottom. Your workspace will then appear as shown in Figure 3.

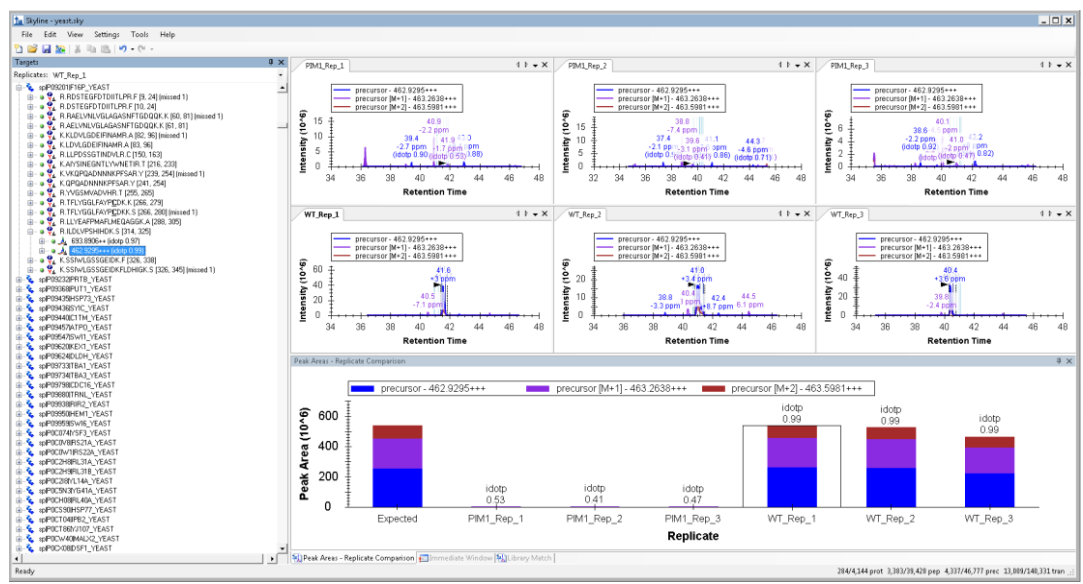

**Figure 3.** Example of viewing an individual peptide's quantification using Skyline. The +3 charge state of the peptide sequence ILDLVPSHIHK from the protein Fructose-1,6-bisphosphatase is essentially absent from the PIM1 mutant, in agreement with the other peptides measured for this protein.

## 4. Expected Results

The procedure presented in this protocol should detect changes in abundance of individual proteins given sufficient replicates are collected to achieve enough statistical power. From the training data provided, this protocol detects 223 protein changes with adjusted p-value < 0.05 (Supplemental Table 1, Figure 4). As described in the initial publication of this data [13], the protein Isu1p is altered (Figure 5). If there are no real changes in the comparison, or the data is too noisy either due to variation introduced during sample processing or data collection, then no changes may be detected. The resulting protein changes can be visualized in skyline as shown in Figure 5, or the all the protein changes can be visualized together using volcano plots as shown in Figure 4 (Supplemental R script). Interpretation of the proteomic changes discovered with this protocol should be done in the context of the perturbation used as described in Veling *et al.* [13].

Because the described analysis is modular, there are additional options for each step of analysis. For example, any other database search algorithm could be used for peptide identification, such as MS-GF+ [14]. Also, because Skyline is used for quantification in this protocol, the peak areas can be output manually with isotope-level granularity and used with any downstream statistical analysis. For example, the peak areas could be output for each sample, filtered for an arbitrary number of top N peptides, and then input to MSstats or mapDIA [15].

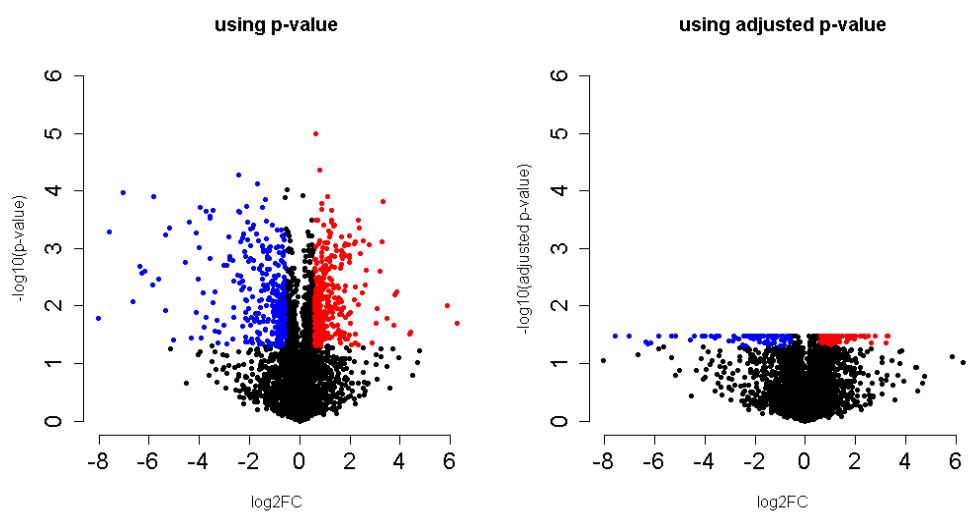

**Figure 4.** Volcano plots from the tutorial data results comparing the proteome of ΔPIM1/WT using MSstats output with either (a) p-value or (b) adjusted p-value. Red indicates an increase and blue indicates a decrease at a cutoff of at least 1.5 fold and p-value or adjusted p-value cutoff < 0.05.

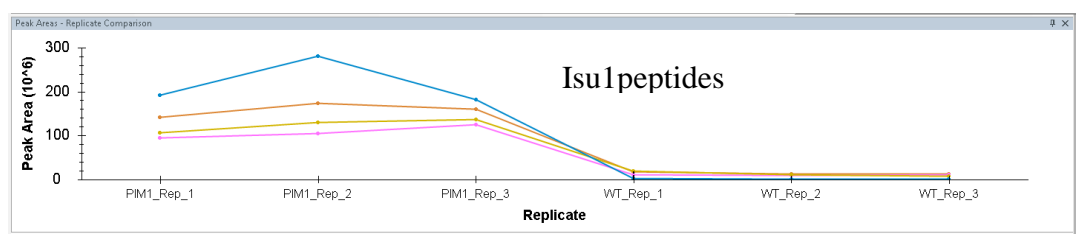

**Figure 5.** Peptide areas measured by skyline for the protein Isu1 showing a concerted increase in the PIM1 knockout.

**Supplemental Information:** The raw data is available from massive.ucsd.edu under the accession MSV000083136 ([ftp://massive.ucsd.edu/MSV000083136/raw/](ftp://massive.ucsd.edu/MSV000083136/raw/)), and the output from MSstats analysis of this tutorial data is available on zenodo [https://doi.org/10.5281/zenodo.1555048](https://doi.org/10.5281/zenodo.1555048).

**Author Contributions:** conceptualization, J.G.M.; methodology, J.G.M.; software, J.G.M.; validation, J.G.M.; formal analysis, J.G.M.; investigation, J.G.M.; resources, J.G.M.; data curation, J.G.M.; writing—original draft preparation, J.G.M.; writing—review and editing, J.G.M.; visualization, J.G.M.; supervision, J.G.M.; project administration, J.G.M.; funding acquisition, J.G.M.

**Funding:** JGM was supported by an NIH T15 fellowship (T15 LM007359).

**Acknowledgments:** None.

**Conflicts of Interest:** The author declares no conflict of interest.

**Appendix A**

This appending contains additional figures for the FragPipe portion of the data analysis that would otherwise disrupt the flow of the protocol.

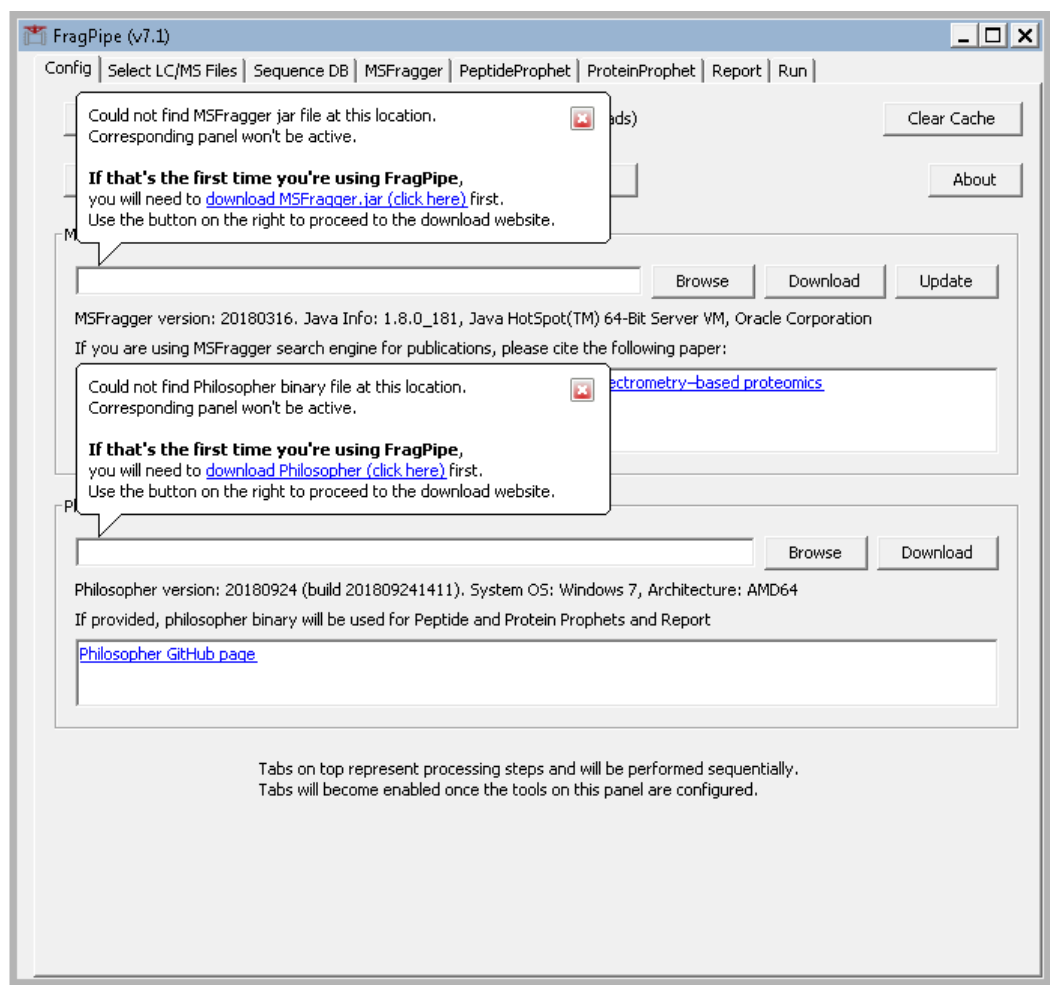

**Figure A1.** Initial FragPipe screen asking for MSFragger.jar and philosopher.exe locations.

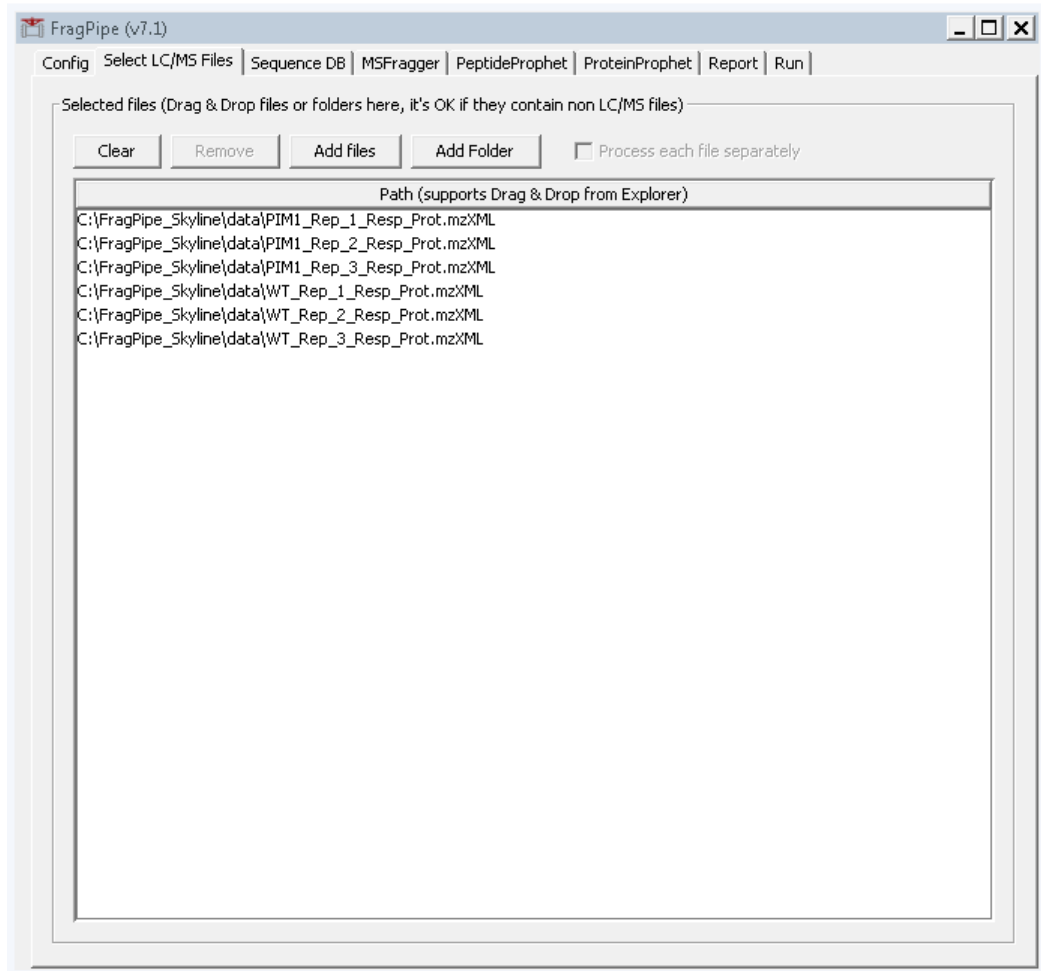

**Figure A2.** FragPipe setup for loading converted mass spectrometry data files.

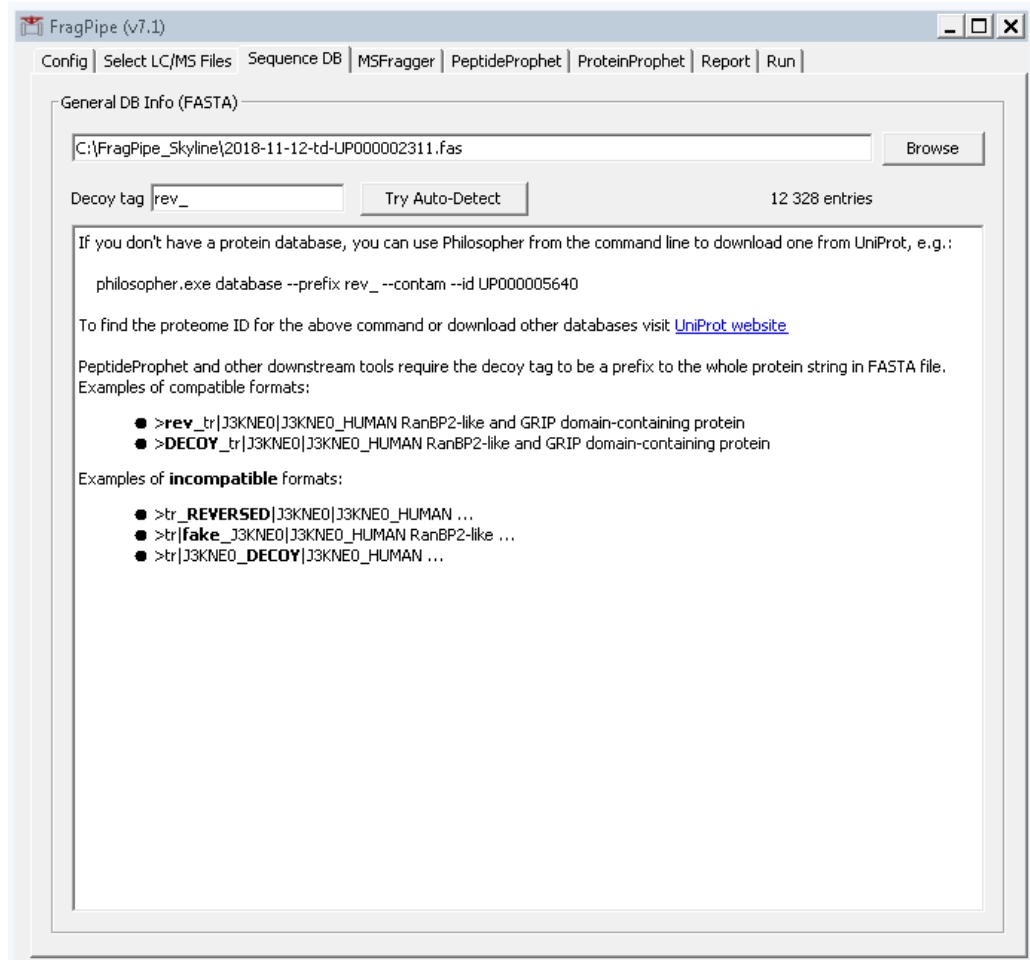

**Figure A3.** FragPipe database entry tab.

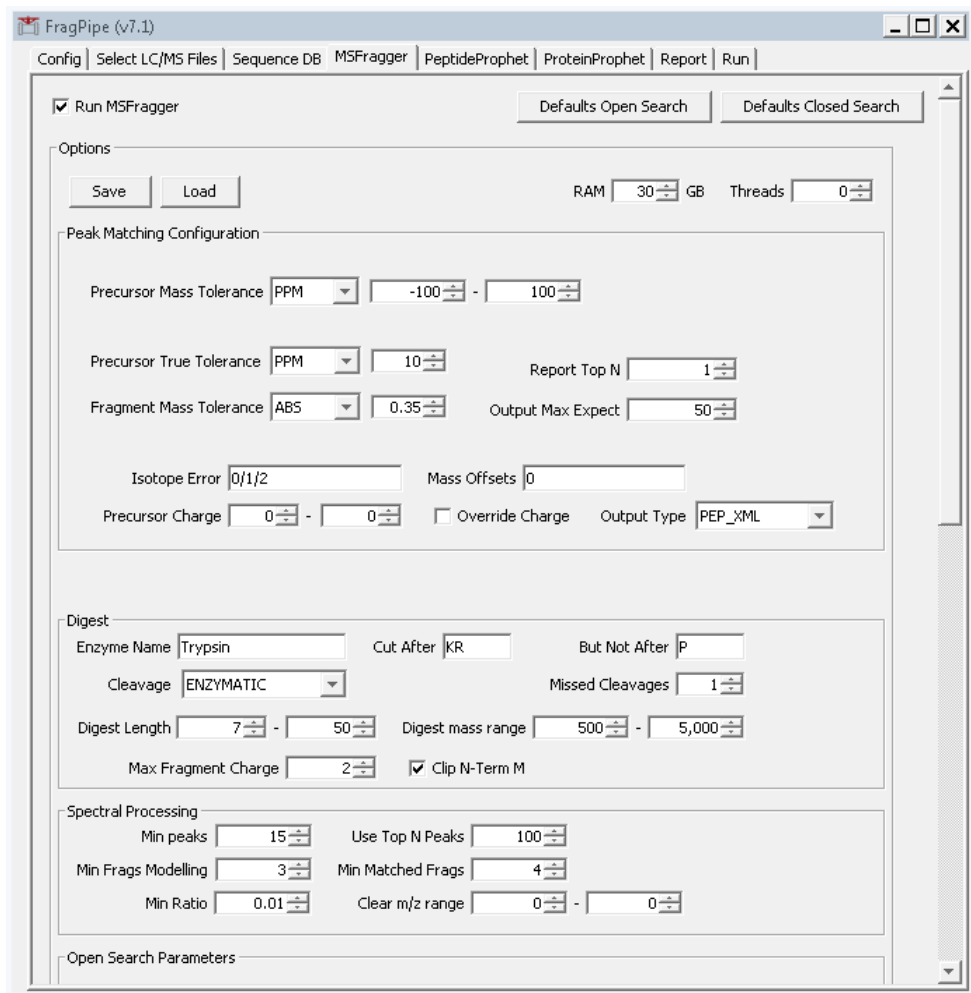

**Figure A4.** FragPipe MSFragger settings tab used for the tutorial data.

**Appendix B**

This appending contains additional figures for the Skyline portion of the data analysis that would otherwise disrupt the flow of the protocol.

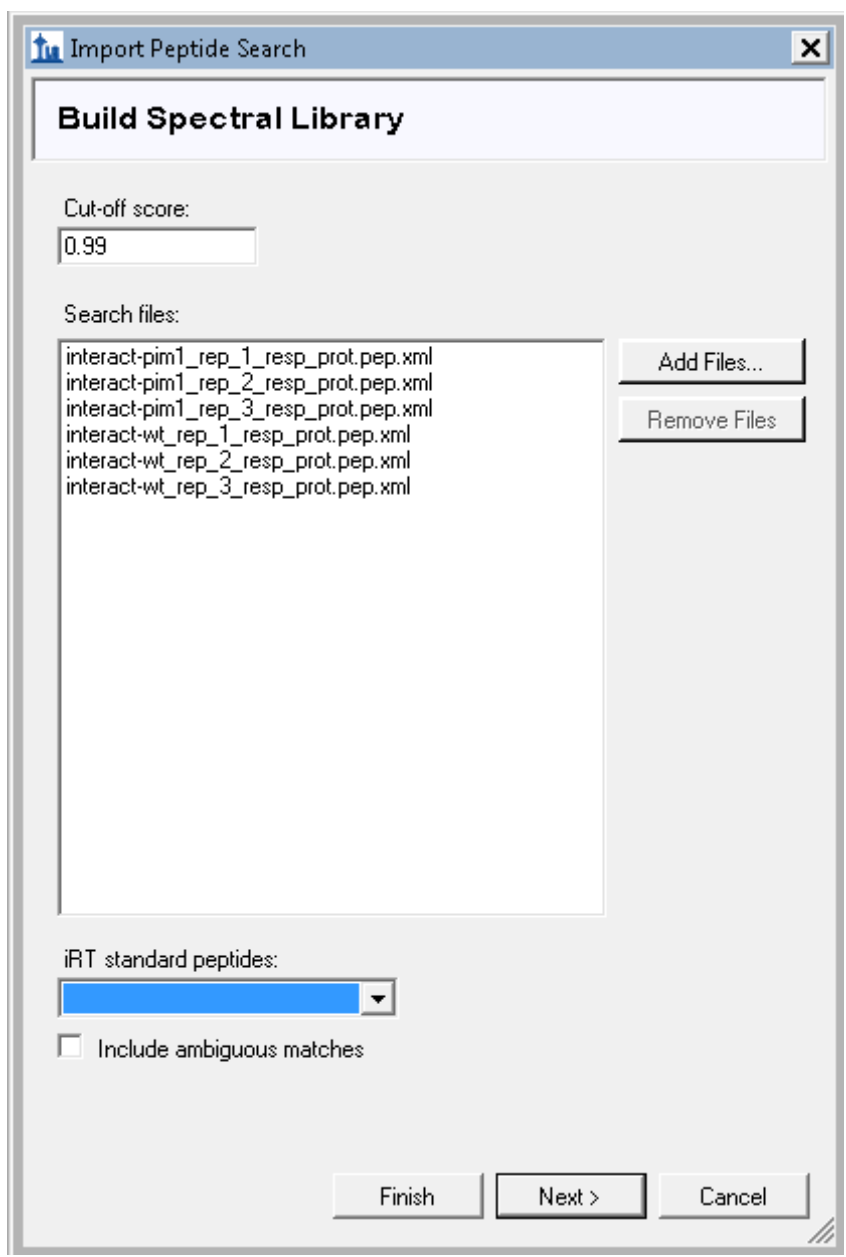

**Figure B1.** First Skyline import wizard screen for selection of the PeptideProphet results.

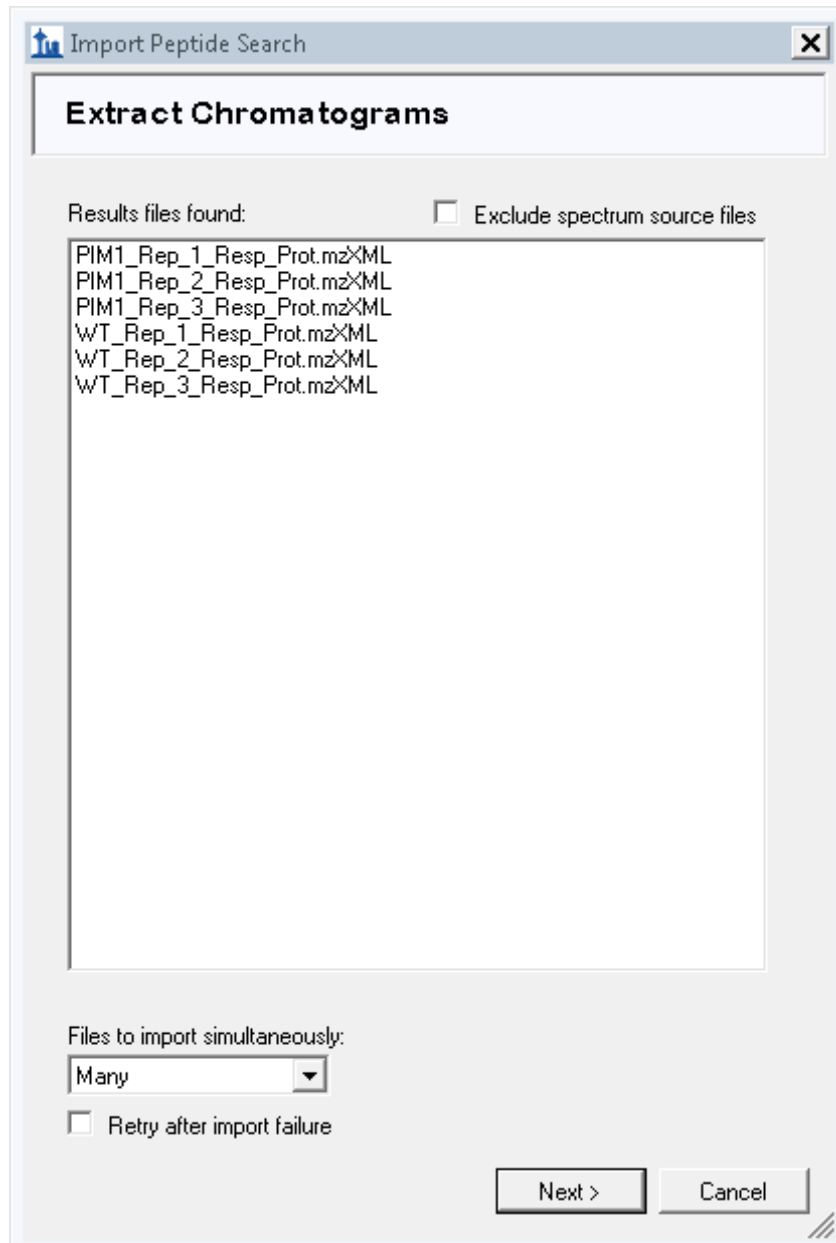

**Figure B2.** Expected view of "Extract Chromatograms" screen during Skyline import wizard.

**Figure B3.** Settings for Orbitrap-measured precursor signal extraction used for the tutorial data.

**Figure B4.** Settings for FASTA file import of the Skyline wizard.

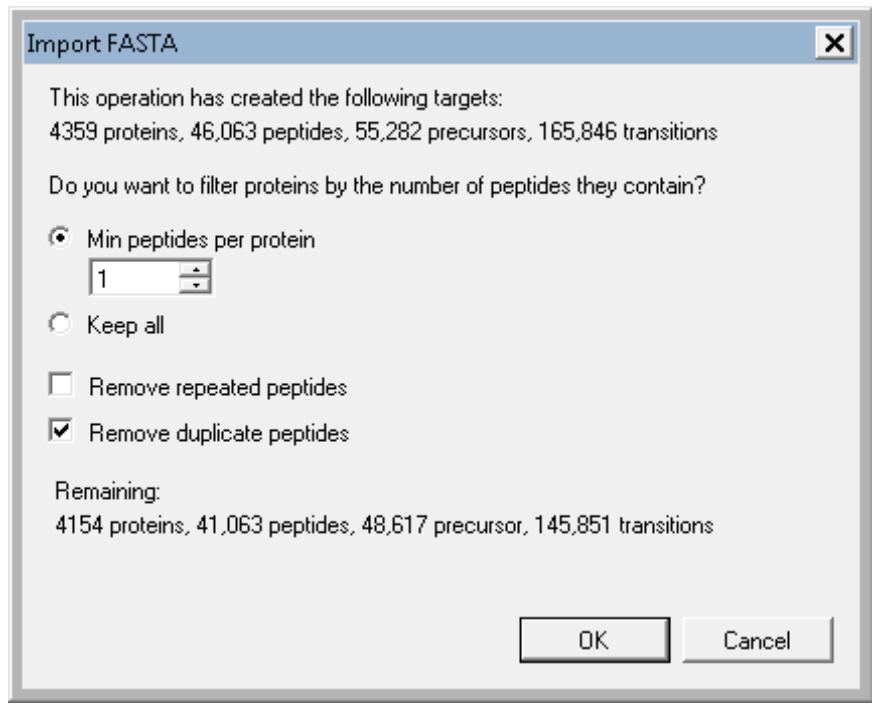

**Figure B5.** Settings for peptide filtering for the tutorial data.

**Figure B6.** MSstats replicate annotation settings for the tutorial data.

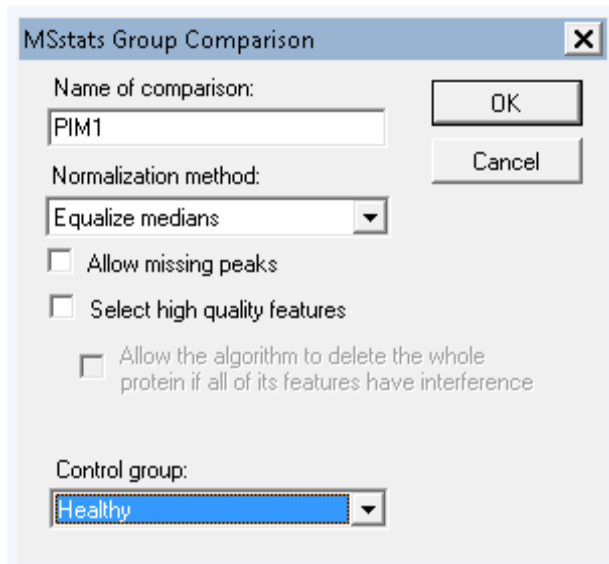

**Figure B7.** MSstats "Group Comparisons" settings for the tutorial dataset.